\documentclass[dvips]{article}

\usepackage{icrc2011}

\date{}

\title{LOPES 3D reconfiguration and first measurements}

\newcommand{\etal}{\MakeLowercase{\textit{et al. }}} 
\shorttitle{D. Huber \etal LOPES 3D}

\authors{
D.~Huber$^{1}$,
W.D.~Apel$^{2}$,
J.C.~Arteaga$^{1,14}$,
L.~B\"ahren$^{3}$,
K.~Bekk$^{2}$,
M.~Bertaina$^{4}$,
P.L.~Biermann$^{5}$,
J.~Bl\"umer$^{1,2}$,
H.~Bozdog$^{2}$,
I.M.~Brancus$^{6}$,
P.~Buchholz$^{7}$,
E.~Cantoni$^{4,8}$,
A.~Chiavassa$^{4}$,
K.~Daumiller$^{2}$,
V.~de~Souza$^{1,15}$,
F.~Di~Pierro$^{4}$,
P.~Doll$^{2}$,
R.~Engel$^{2}$,
H.~Falcke$^{3,9,5}$,
M. Finger$^{1}$, 
B.~Fuchs$^{1}$,
D.~Fuhrmann$^{10}$,
H.~Gemmeke$^{11}$,
C.~Grupen$^{7}$,
A.~Haungs$^{2}$,
D.~Heck$^{2}$,
J.R.~H\"orandel$^{3}$,
A.~Horneffer$^{5}$,
T.~Huege$^{2}$,
P.G.~Isar$^{2,16}$,
K.-H.~Kampert$^{10}$,
D.~Kang$^{1}$, 
O.~Kr\"omer$^{11}$,
J.~Kuijpers$^{3}$,
K.~Link$^{1}$, 
P.~{\L}uczak$^{12}$,
M.~Ludwig$^{1}$,
H.J.~Mathes$^{2}$,
M.~Melissas$^{1}$,
C.~Morello$^{8}$,
J.~Oehlschl\"ager$^{2}$,
N.~Palmieri$^{1}$,
T.~Pierog$^{2}$,
J.~Rautenberg$^{10}$,
H.~Rebel$^{2}$,
M.~Roth$^{2}$,
C.~R\"uhle$^{11}$,
A.~Saftoiu$^{6}$,
H.~Schieler$^{2}$,
A.~Schmidt$^{11}$,
F.G.~Schr\"oder$^{2}$,
O.~Sima$^{13}$,
G.~Toma$^{6}$,
G.C.~Trinchero$^{8}$,
A.~Weindl$^{2}$,
J.~Wochele$^{2}$,
M.~Wommer$^{2}$,
J.~Zabierowski$^{12}$,
J.A.~Zensus$^{5}$
}
\afiliations{
$^1$ Institut f\"ur Experimentelle Kernphysik, KIT - Karlsruher Institut f\"ur Technologie, Germany\\
$^2$ Institut f\"ur Kernphysik, KIT - Karlsruher Institut f\"ur Technologie, Germany\\
$^3$ Radboud University Nijmegen, Department of Astrophysics, The Netherlands\\
$^4$ Dipartimento di Fisica Generale dell' Universit\`a Torino, Italy\\
$^5$ Max-Planck-Institut f\"ur Radioastronomie Bonn, Germany\\
$^6$ National Institute of Physics and Nuclear Engineering, Bucharest, Romania\\
$^7$ Fachbereich Physik, Universit\"at Siegen, Germany\\
$^8$ Istituto di Fisica dello Spazio Interplanetario, INAF Torino, Italy\\
$^9$ ASTRON, Dwingeloo, The Netherlands\\
$^{10}$ Fachbereich Physik, Universit\"at Wuppertal, Germany\\
$^{11}$ Institut f\"ur Prozessdatenverarbeitung und Elektronik, KIT - Karlsruher Institut f\"ur Technologie, Germany\\
$^{12}$ Soltan Institute for Nuclear Studies, Lodz, Poland\\
$^{13}$ Department of Physics, University of Bucharest, Bucharest, Romania\\
\scriptsize{
$^{14}$ now at: Univ Michoacana, Morelia, Mexico;
$^{15}$ now at: Univ S$\tilde{a}$o Paulo, Inst. de F\'{\i}sica de S\~ao Carlos, Brasil; 
$^{16}$ now at: Inst. Space Sciences, Bucharest, Romania 
}
}

\email{daniel.huber@kit.edu}

\abstract{The Radio detection technique of high-energy cosmic rays is based on the radio signal emitted by the charged particles in an air shower due to their deflection in the Earth's magnetic field. The LOPES experiment at Karlsruhe Institute of Technology, Germany with its simple dipoles made major contributions to the revival of this technique. LOPES is working in the frequency range from $40$ to $80$\,MHz and was reconfigured several times to improve and further develop the radio detection technique. In the current setup LOPES consists of 10 tripole antennas which measure the complete electric field vector of the radio emission from cosmic rays. LOPES is the first experiment measuring all three vectorial components at once and thereby gaining the full information about the electric field vector and not only a two-dimensional projection. Such a setup including also measurements of the vertical electric field component is expected to increase the sensitivity to inclined showers and help to advance the understanding of the emission mechanism. We present the reconfiguration and calibration procedure of LOPES 3D and discuss first measurements.}

\keywords{ Radio detection LOPES calibration }

\begin{document}
\maketitle
\section{Introduction}
The observation of cosmic rays with energies $\geq10^{17}$\,eV is a challenging field of research. For the highest energies one has to go to extremely large detector arrays on the ground and to face several difficulties in the reconstruction and the measurement itself. To gain the best results, different observation techniques have to be combined. Radio detection is a promising candidate to observe the air shower development, because all charged particles, especially $e^{-}$ and $e^{+}$, get deflected in the Earth's magnetic field and emit a radio signal \cite{Haungsradio}. The radio emission is not absorbed in the atmosphere and is only influenced by strong atmospheric electric fields such as those present during thunderstorms \cite{BuitinkApelAsch2006}. It is so far the only detection technique that combines a long up time and sensitivity to the shower development because, in contrast to fluorescence measurements, it can also measure at daytime \cite{haungs2003}. Therefore it is desirable to fully exploit the potential of this detection technique. One advancement within this detection method is the vectorial measurement of radio emission. To study the feasibility of these vectorial measurements, the LOPES experiment \cite{FalckeNature2005} was reconfigured to be now able to measure all three components of the electric field vector from the radio emission of cosmic ray induced air showers and not only a two-dimensional projection. With the complete measured electric field vector a better comparison between different emission mechanisms is possible, the sensitivity to inclined air showers is increased and the direction reconstruction can be crosschecked and improved. The achieved results have the potential to impact future experiments and analyses. LOPES once more acts as research and development array for large scale applications.  

\section{Reconfiguration}

\subsection{Hardware}
With the reconfiguration of LOPES to LOPES 3D some hardware changes had to be made e.g. the antenna type was changed which made new preamplifier and new cabling necessary. The new hardware is discussed in detail in the following paragraphs.

 \begin{figure}[h]
  \vspace{5mm}
  \centering
  \includegraphics[width=2.3in]{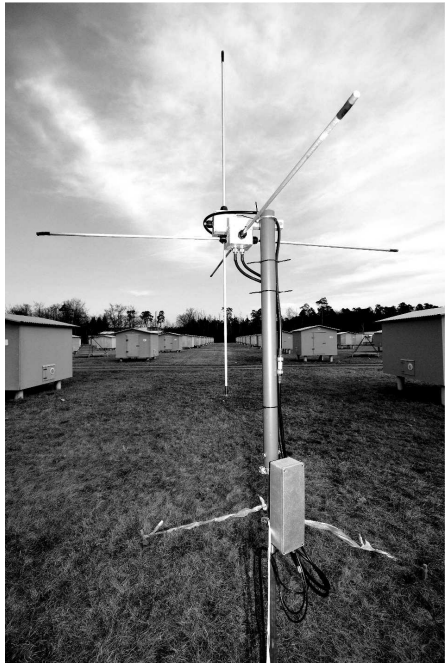}
  \caption{Picture of the tripole, the new antenna type used at LOPES 3D. The metal box at the pole is the housing for the preamplifier.}
  \label{figure:2}
 \end{figure}
 
\subsubsection{Antenna type}
The antenna type used for LOPES 3D is a tripole, see figure \ref{figure:2}. This antenna consists of three crossed dipoles and covers one channel per dipole, thus three channels per antenna. The tripole provides high sensitivity, a homogeneous setup  and symmetrical characteristics for each channel. Having homogeneous and symmetrical antenna characteristics is essential for vectorial analyses and reduces the influence of systematic uncertainties.

\subsubsection{Preamplifier}
The low noise amplifiers (LNA) used for LOPES 3D were originally designed for the Auger Engineering Radio Array AERA \cite{Huege 2010}. They provide low power consumption, high stability and a very low noise level. Because they are two channel amplifiers one needs two amplifiers per tripole which results in one spare amplifier channel per tripole.

\subsubsection{Beacon}
\label{beacon}
The beacon is a reference emitter emitting sine waves. It is needed to improve and monitor the timing of the experiment \cite{SchroederTimeCalibration2010}. With the beacon, a timing accuracy of $1$\,ns, which is needed for digital interferometry, is achieved. The phase differences of the signals from the beacon are constant due to geometrical reasons. With the known phase differences from the calibration and the actual measured ones the timing can be monitored and if necessary corrected.  In order to be also seen in the new, vertical polarization, the beacon had to be rotated and the emission power was increased. 

\subsection{Antenna positions}
LOPES 3D has ten antenna positions, since the LOPES data acquisition provides 30 channels and each tripole allocates three channels. For the new positions several requirements had to be taken into account:
\begin{itemize}
\item Cover a huge area.
\item Avoid regularities since they lead to higher side lobes \cite{Almamemo} when using LOPES 3D as digital radio interferometer.
\item Reuse of cables from the previous setup of LOPES.
\item Reduce the effort and the danger of damaging the hardware during the reconfiguration by choosing antenna positions where old LOPES antennas were standing.
\end{itemize}

\begin{figure}[h]
  \vspace{5mm}
  \centering
  \includegraphics[width=3in]{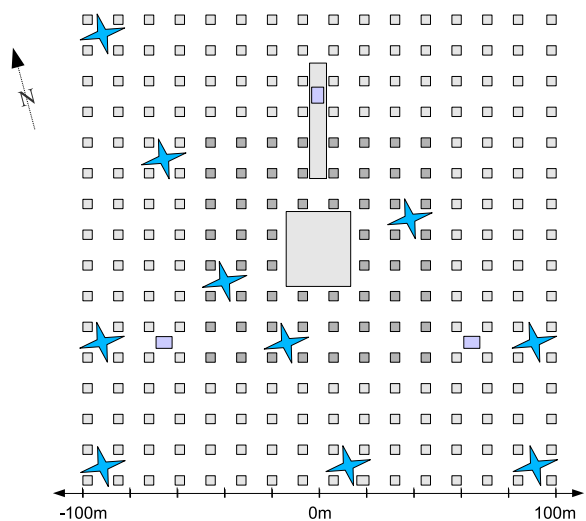}
  \caption{Scheme of the LOPES 3D antenna positions within the KASCADE \cite{kascade} array. The LOPES 3D antennas are marked as stars, the KASCADE detector stations as gray squares and the KASCADE-Grande detectors are shown as gray rectangles.}
  \label{figure:1}
 \end{figure}

The resulting layout can be seen in figure \ref{figure:1}. LOPES 3D measures in this configuration since May 2010.

\section{Calibration}
After the reconfiguration, a complete calibration needed to be done. The single steps are explained in detail in the following. 

\subsection{Measurement of the antenna positions}
The requirement in the timing accuracy when using LOPES 3D as digital radio interferometer is $1$\,ns.
This accuracy in the timing can be converted to an accuracy in the position by dividing through the speed of light and is for LOPES $\frac{1\,ns}{3 \times 10^{8}\frac{m}{s}}=30$\,cm. The antenna positions of LOPES are measured with a differential GPS system. This system has a resolution of $\approx 1.5$\,cm in East and West and  $\approx 2$\,cm in the height which clearly exceeds the requirements.

\subsection{Simulation of the antenna characteristics}
\begin{figure}[h]
  \vspace{5mm}
  \centering
  \includegraphics[width=2.8in]{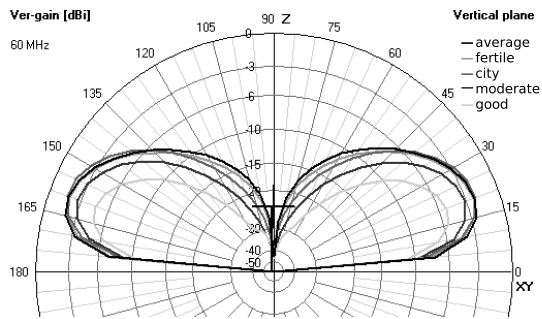}
  \caption{Simulated gain pattern AT $60$\,MHz for the vertical dipole for different ground conditions.}
  \label{figure:8}
 \end{figure}

To determine the characteristics of an antenna there are two possibilities. On the one hand one can try to measure the characteristics, on the other hand one can use a simulation to calculate the antenna behavior. In the case of LOPES 3D the antenna characteristics were simulated. For this simulation the latest public available version of the numerical electromagnetic code NEC2 \cite{4NEC2} was used. With these simulations the gain pattern as well as the influences of the different ground types on the antenna characteristics can be studied, see also figure \ref{figure:8}. For LOPES the 4NEC2 average ground was chosen since this ground is between the city ground (KASCADE detector huts) and fertile land (grassland around the huts) which describe the LOPES site. 

\subsection{Measurements of the electronics' delay}
\begin{figure}[h]
  \vspace{5mm}
  \centering
  \includegraphics[width=4.8in]{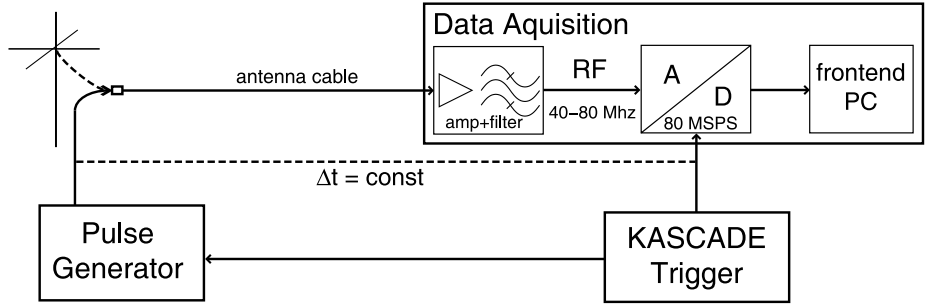}
  \caption{Scheme of the measurement setup for the delay calibration. The antenna is drawn to mark the place where it would be connected \cite{SchroederTimeCalibration2010}.}
  \label{figure:4}
 \end{figure}

When LOPES is used as a digital radio interferometer a precise timing is indispensable. In the frequency range of LOPES, $40$ to $80\,$MHz, the relative timing between two channels needs to be known with an accuracy better than $1$\,ns \cite{SchroederTimeCalibration2010}. To gain this high accuracy two steps are necessary. First to get the rough timing by sending test pulses and measuring the time they need to be recorded from the data acquisition. A sketch of the measurement setup is shown in figure \ref{figure:4}. A pulse generator is connected to the antenna cable instead of the antenna. The pulse generator and the data acquisition are both triggered by KASCADE-Grande \cite{kascade-grande}, since only the relative timing between the channels is of interest the time $\Delta t$ does not need to be known.

\begin{figure}[h]
  \vspace{5mm}
  \centering
  \includegraphics[width=3in]{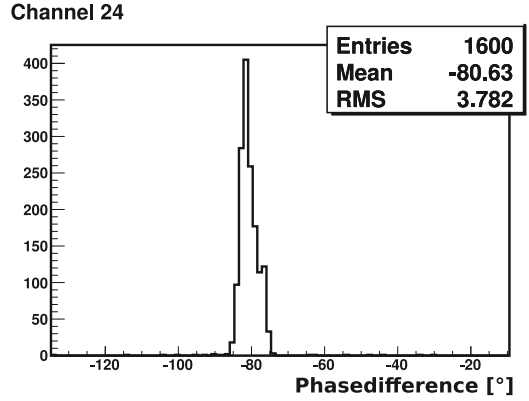}
  \caption{Measured phase differences of 1600 events ($\approx$ half a day) between channel 1 and 24 for $68.1$\,MHz. The RMS of $3.782$ corresponds to a delay of $0.15$\,ns.}
  \label{figure:5}
 \end{figure}

Second step is to do the fine tuning by measuring the reference phases of the signal from the beacon. With the method explained in \ref{beacon} a timing accuracy of $\leq1$\,ns is achieved. An example for the distribution of the phase differences between channel 24 and 1 is shown in figure \ref{figure:5}.

\subsection{Absolute amplitude calibration}
\begin{figure}[h]
  \vspace{5mm}
  \centering
  \includegraphics[width=3.in]{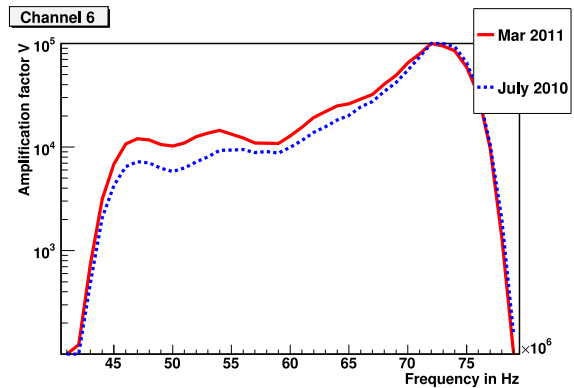}
  \caption{Amplification factors of channel 6 measured at two different days. The different amplification factors originate from the uncertainties in the measurement, from environmental conditions and from the aging of the electronics.}
  \label{figure:3}
 \end{figure}

In order to know which field strength at the antenna corresponds to which analog to digital converter (ADC) value, the whole signal chain needs to be calibrated with a known reference source. The reference source needs to be arranged over the antenna of the channel to be calibrated. It is important to be in the far field  which means that in  case of LOPES the distance between antenna and reference source should be at least $10\,$m \cite{NehlsHakenjosArts2007}. The position of the reference source is measured with differential GPS and the orientation is checked by eye. The requirements are less than $7$\,$^{\circ}$ deviation in the alignment and less than $0.5\,$m deviation from the position. With the known power at the antenna and the measured ADC values the frequency-dependent calibration factors for the amplification can be calculated. An example for two amplification factors of the same channel measured at different days is shown in figure \ref{figure:3}. Using this method one gains the advantage of being able to calibrate the complete analog signal chain, which reduces systematic uncertainties and gives the possibility to detect potential interferences between the different hardware components.

\section{Performance}
After the reconfiguration and calibration of LOPES 3D also the performance needed to be checked. For that purpose some simple analyses were done. To declare an event as detected, the signal-to-noise ratio of the cross-correlation beam was required to exceed a value of 6 for KASCADE-Grande triggered and 8 for KASCADE triggered events. The cross-correlation beam is calculated when using LOPES 3D as digital radio interferometer \cite{HornefferThesis2006,katrin}. It gives information on the coherent part of a radio signal and increases significantly the signal-to-noise ratio of a radio pulse. In a noisy environment like the KIT the cross-correlation beam is the first step for all further analyses. After these cuts the average event rate of LOPES 3D is $\approx 1.75$\,$\frac{events}{week}$ which is well within the expectations of a factor of $\frac{1}{3}$ fewer antenna positions since with LOPES 30 an average event rate of $\approx 3.5$\,$\frac{events}{week}$  was recorded \cite{NehlsThesis2008}. The detection efficiency scales with the square root of the number of antennas and linear with the primary energy. With the known index of the integrated cosmic ray flux spectrum $2$ the event rate can be determined to scale linear with the number of antennas. The average event rate of LOPES 3D was calculated for the dataset available from 2010.

\section{Conclusions}
The reconfiguration of LOPES to LOPES 3D regarding all requirements has been successful. Now LOPES 3D is the first experiment that measures all three components of the electric field vector from cosmic ray induced air shower radio emission.  The experiment is calibrated and fully operational. With LOPES 3D it is possible to study the feasibility of vectorial polarization measurements for the radio detection technique. First analyses confirmed the performance as expected. In the future, sophisticated vectorial analyses need to be done to test the potential of vectorial radio measurements. 


\end{document}